\documentclass[12pt]{article}
\usepackage{cite,graphicx}

\newcommand{\newc}{\newcommand}
\newc{\gsim}{\lower.7ex\hbox{$\;\stackrel{\textstyle>}{\sim}\;$}}
\newc{\lsim}{\lower.7ex\hbox{$\;\stackrel{\textstyle<}{\sim}\;$}}
\newc{\gev}{\,{\rm GeV}}
\newc{\mev}{\,{\rm MeV}}
\newc{\ev}{\,{\rm eV}}
\newc{\kev}{\,{\rm keV}}
\newc{\tev}{\,{\rm TeV}}

\newc{\mz}{M_Z}
\newc{\mpl}{M_*}
\newc{\mw}{m_{\rm weak}}
\renewcommand{\epsilon}{\varepsilon}
\def\beq{\begin{equation}}
\def\eeq{\end{equation}}
\def\bea{\begin{eqnarray}}
\def\eea{\end{eqnarray}}
\newc{\ie}{{\it i.e.}}          \newc{\etal}{{\it et al.}}
\newc{\eg}{{\it e.g.}}          \newc{\etc}{{\it etc.}}
\newc{\cf}{{\it c.f.}}
\def\bar#1{\overline{#1}}
\def\vev#1{\left\langle #1 \right\rangle}

\def\inv{^{\raise.15ex\hbox{${\scriptscriptstyle -}$}\kern-.05em 1}}
\def\lbar{{\lower.35ex\hbox{$\mathchar'26$}\mkern-10mu\lambda}} 

\def\to{\rightarrow}

\let\p=\partial
\let\<=\langle
\let\>=\rangle

\let\+=\uparrow

\let\de=\delta

\let\ep=\varepsilon

\let\la=\lambda

\let\th=\theta

\begin{document}
\thispagestyle{empty}
\vspace*{.5cm}
\noindent
\hspace*{\fill}{CERN-TH/2002-101}\\
\vspace*{2.5cm}

\begin{center}
{\Large\bf The Flavour Hierarchy and See-Saw Neutrinos\\[.3cm] 
from Bulk Masses in 5d Orbifold GUTs}
\\[2.5cm]
{\large Arthur Hebecker and John March-Russell}\\[.5cm]
{\it Theory Division, CERN, CH-1211 Geneva 23, Switzerland}
\\[.2cm]
(May 14, 2002)
\\[1.1cm]

{\bf Abstract}\end{center}
\noindent
In supersymmetric grand unified theories (GUTs) based on
$S^1/(Z_2\times Z'_2)$ orbifold constructions in 5 dimensions, Standard
Model (SM) matter and Higgs fields can be realized in terms of 5d
hypermultiplets.  These hypermultiplets can naturally have
large bulk masses, leading to a localization of the zero modes at one of
the two branes or to an exponential suppression of the mass of the
lowest-lying non-zero mode.  We demonstrate that these dynamical features
allow for the construction of an elegant 3-generation SU(5) model in
5 dimensions that explains all the hierarchies between fermion masses and
CKM matrix elements in geometrical terms.  Moreover, if U(1)$_\chi$ (where
SU(5)$\times$U(1)$_\chi \subset$~SO(10)) is gauged in the bulk, but broken
by the orbifold action at the SM brane, the right-handed neutrino mass
scale is naturally suppressed relative to $M_{\rm GUT}$.  Together with our
construction in the charged fermion sector this leads, via the usual
see-saw mechanism, to a realistic light neutrino mass scale and large
neutrino mixing angles.
 
\newpage

\setcounter{page}{1}

\section{Introduction}

Supersymmetric (SUSY) grand unification provides an elegant explanation of 
the fermion quantum numbers and of the relative strength of the three gauge
couplings of the standard model (SM).  However, conventional 4d GUTs
possess less attractive features such as a complicated GUT-scale Higgs sector,
unsuccessful first and second generation analogues of the $m_b/m_\tau$
mass prediction, and dangerous dimension-5 proton decay operators
arising from Higgsino exchange. 

Recently, an elegant solution to these problems has been proposed in the 
context of SU(5)~\cite{kaw,AF,HN,HMR,HMR2,su5} and SO(10)~\cite{so10} 
unification. The GUT gauge symmetry is now realized in 5 or more space-time 
dimensions and broken to the SM group by compactification on an orbifold, 
utilizing boundary conditions that violate the GUT-symmetry. In the most 
studied case of 5 dimensions both the GUT group and 5d supersymmetry are 
broken by compactification on $S^1/(Z_2\times Z_2')$, leading to an N=1 SUSY
model with SM gauge group.  This construction provides elegant solutions to
the problems of conventional GUTs with Higgs breaking, including 
doublet-triplet splitting, dimension-5 proton decay, and Yukawa 
unification in the first two generations, while maintaining, at least at 
leading order, the desired gauge coupling unification~\cite{HN,HMR,CPRT}.

Although the hierarchy between the strong coupling scale $M$ of the 5d gauge 
theory and the compactification scale $1/R$ $(MR\sim 10^2\cdots 10^3)$ can be 
used to generate a fermion mass hierarchy~\cite{HMROS,hns}, the construction 
of a realistic three-generation model in 5 dimensions proves difficult (see 
also~\cite{orbf} and, in particular,~\cite{hn2}, where problems very similar 
to the present investigation have been attacked with different tools). 
Furthermore, the slight discrepancy between the unification scale and the 
phenomenologically preferred right-handed (rhd) neutrino mass scale in see-saw 
models is generically enhanced in orbifold GUTs.  In this letter, we
demonstrate that 5d bulk masses, which are naturally present in the theories
under consideration, allow for the construction of realistic models with the 
correct fermion mass hierarchy and rhd neutrino scale.  The main ingredients 
we employ are the bulk-mass-driven localization of the zero mode and, in 
the absence of a zero-mode, the exponential suppression of the effective 
4d mass in the limit where the bulk mass term becomes large.

\section{Bulk masses in 5d SUSY}

Recall first that the Lagrangian for a 5d hypermultiplet (written in terms 
of two 4d chiral superfields $H$ and $H^c$~\cite{agw}) is
\beq
{\cal L}=\int_{\theta^2\bar{\theta}^2}\left(H^\dagger H+H^cH^{c\dagger}
\right)+\left(\int_{\theta^2}H^c\partial_5H+\mbox{h.c.}\right)\, .
\eeq
In the general case of a 5d orbifold, the fundamental region in the 
direction of $x^5=y$ is an interval, bounded by two inequivalent fixed 
points at $y=0$ and $y=l$ (where $l=\pi R/2$ in the case of $S^1/(Z_2\times 
Z_2')$). Each of the fixed points is invariant under a $Z_2$ reflection of 
the original 5d theory. The two superfields ($H,H^c$) of a singlet 
hypermultiplet in the bulk necessarily have opposite parities under each
of the $Z_2$'s.  
The above Lagrangian can be supplemented by the 5d Lorentz invariant
mass terms
\beq
{\cal L}_{\rm mass}=m_o\left(\int_{\theta^2}H^cH+\mbox{h.c.}\right)
+\frac{1}{2}m_e\left(\int_{\theta^2}\left\{H^2+(H^c)^2\right\}+\mbox{h.c.}
\right)\,,
\eeq
which are respectively odd and even under parity.  We will mainly be 
interested in the case of gauged hypermultiplets, where the even mass term 
is forbidden ($m_e=0$, $m_o=m$). Furthermore, we will not be concerned with
the dynamical realization of the odd mass term (see, e.g.,~\cite{fi}), but
simply include it into our effective field theory Lagrangian as a leading 
operator consistent with all the symmetries of the model. 

The localization of fermionic zero modes in the presence of $y$-dependent 
mass terms is a familiar phenomenon~\cite{jr} which has been used in the 
context of model building in extra dimensions by many authors (see, 
e.g.,~\cite{as,bm}). Thus, we can content ourselves here with recalling the 
basic features relevant for 5d orbifold constructions.~\footnote{
We would like to emphasize that, in contrast to~\cite{as} and many related 
papers, we do not place Gaussian zero modes at various points in the bulk 
but restrict ourselves to mass terms that are constant between $y=0$ and $l$
and only allow for a peaking of the modes at either boundary.}

The equation for the
$y$-dependent profile $H(y)$ in the presence of the odd mass $m$ is 
\beq
\biggl( \p_y^2 - m^2 + m_4^2 + 
2m\left( \de(y) - \de(y-l)\right) \biggr) H(y) =0\,,
\eeq
where $m_4$ is the effective mass of the mode in 4d, and the $\delta$ 
function terms arise from the discontinuity of the odd mass function at the 
fixed points. (A similar equation holds for $H^c(y)$.) 

The superfields $H$ and $H^c$ can have either the same or opposite parities 
at $y=0,l$. To discuss the first case, assume that $H$ and $H^c$ have the 
parities $(+,+)$ and $(-,-)$ at the two boundaries.  A simple analysis of 
the equation in this case shows that $H$ has a zero-mode with bulk profile 
\beq
H(y) =  e^{-ym} 
\label{zm}
\eeq
while all other KK masses, including those of $H^c$, are ${\cal O}(m)$
or larger. (Since $m$ is naturally of the order of the UV scale $M$, we 
assume $m\gg 1/R$ in our analysis.)  Thus, depending on the sign of $m$, 
the zero mode is exponentially localized at the left ($m>0$) or right 
($m<0$) boundary of the orbifold. 

In the second case, we choose $H$ and $H^c$ to have the parities $(+,-)$ 
and $(-,+)$ at the two boundaries.  Although now no zero-mode exists, two 
4d superfield excitations are found to have a mass much smaller than $m$
(for $m>0$). They can be characterised as an $H$ and an $H^c$ mode with bulk 
profiles 
\beq
H(y)\simeq \left( e^{-ym}-e^{(y-2l)m} \right)~~\mbox{and}~~
H^c(y)\simeq \left( e^{(y-l)m}-e^{-(y+l)m} \right)\,,
\label{bp}
\eeq
which are linked by a 4d Dirac-type mass 
\beq
m_4 \simeq 2 m e^{-ml} \, .
\label{m4}
\eeq
(for canonical 4d superfield normalization of the kinetic term).  As
can be seen from Eq.~(\ref{bp}), the $H$ and $H^c$ mode are
exponentially localized at the two opposite boundaries of the orbifold.

\section{Bulk masses in orbifold GUTs}\label{orbbm}

We begin by recalling the basic structure of the Kawamura
model~\cite{kaw}, which is based on a 5d super Yang-Mills theory on 
$I\!\!R^4\times S^1$, where the $S^1$ is parameterised by $y\in[0,2\pi R)$. 
The field space is then restricted by imposing the two discrete $Z_2$ 
symmetries, $y\to -y$ and $y'\to -y'$ (with $y'=y-\pi R/2$). The action of 
the $Z_2$'s in field space is specified by the two gauge twists $P$ and 
$P'$. If the original gauge group is SU(5) and the gauge twists are chosen 
as $P=1$ and $P'=$ diag$(1,1,1,-1,-1)$, the full SU(5) gauge symmetry exists
in the bulk and on the SU(5) brane at $y=0$, while at $y=l$ only the SM gauge
symmetry exists.  As a result, the effective low energy theory is invariant
under only the SM gauge symmetry.

To be specific, in this letter we take the compactification scale to be 
$M_c=1/R \sim 10^{15}$ GeV, slightly lower than the usual GUT scale 
$M_{\rm GUT}\sim 10^{16}$ GeV, while the UV or cutoff scale $M \sim 
10^{17}$ GeV is slightly higher. This situation is generic for the following 
reasons. On the one hand, the mild separation between $M_c$ and $M$ ensures 
that corrections to gauge coupling unification from brane-localized 
operators are under control. At the same time, this allows for a certain 
validity range of the 5d field theory. On the other hand, the `differential 
running' of gauge couplings between $M_c$ and $M$ is somewhat slower than 
the MSSM running below $M_c$~\cite{HN,HMR,nsw,CPRT}, implying 
$M>M_{\rm GUT}$. Moreover, $M$ should not be larger than the scale at which
the 5d gauge theory becomes non-perturbative, $M\lsim (12\pi/\alpha_{\rm 
GUT})M_c$~\cite{HN,HMR}. As we will see in Sect.~\ref{fla}, our flavour 
scenario favours a value $Ml\simeq 300$, which is comfortably within the 
range set by the above restrictions. 

The up- and down-type Higgs fields can be introduced as two hypermultiplets
$(H_u,H_u^c)$ and $(H_d,H_d^c)$ in the bulk, transforming
as a $({\bf 5},\bf{\bar{5}})$ and $({\bf \bar{5}},{\bf 5})$.
After appropriate parity assignment, two doublet zero modes 
emerge. This is the celebrated solution of the doublet-triplet splitting 
problem~\cite{kaw}. Alternatively, the two required doublets can directly 
be introduced on the SM brane, where full SU(5) multiplets are not 
required~\cite{HMR}. In this context, bulk masses can have important 
effects. Firstly, they allow for an exponential localization of the doublet 
zero modes at either the SU(5) or SM brane. Since the 5d SUSY forbids bulk
Yukawa interactions, the SM Yukawa couplings are always brane operators.
This implies that localization can be used for the generation of large fermion
mass hierarchies while keeping the dimensionless coefficients of all 
relevant operators ${\cal O}(1)$.  Secondly, bulk masses allow for the 
interpretation of doublets living on the SM brane as the zero modes of
bulk fields with a large mass. 

Fermion fields can be introduced on the SU(5) brane~\cite{kaw,AF,HN}, on 
the SM brane~\cite{HMR}, or in the bulk~\cite{HN,HMR}. Again, bulk masses 
allow for an interpretation of the brane-localized states as 
limiting cases of the model with bulk fields.  To see this in more detail,
recall that to realize a full ${\bf\bar{5}}$ of SU(5) in the bulk, one 
starts with two hypermultiplets $(\bar{F},\bar{F}^c)$ and $(\bar{F}', 
\bar{F}'^c)$ and chooses parities such that, say, the ${\bf\bar{3}}$ from 
$\bar{F}$ and the ${\bf\bar{2}}$ from $\bar{F}'$ have a zero-mode. Clearly, 
introducing appropriate bulk masses for the two original hypermultiplets, 
these zero modes can now be localized at either of the two branes. 
Similarly, two $\bf 10$ hypermultiplets $(T,T^c)$ and $(T',T'^c)$ in the 
bulk realize the particle content of a full $\bf 10$ as zero modes, which 
can then be localized at either brane. 

The above two paragraphs call for a number of further comments.  Firstly, it 
is now apparent that the `minimal model' of~\cite{HMR}, i.e., a model with
only the gauge sector in the bulk and all other fields on the SM brane, 
can be viewed as a large-bulk-mass limit of a model with bulk fields 
only. In particular, this allows for an understanding of the quantum 
numbers of SM brane fermions in terms of SU(5) representations - an 
important GUT prediction that was previously missing in the minimal model 
of~\cite{HMR}. Thus, the introduction of bulk masses puts the minimal
model, which is phenomenologically attractive because of its simplicity and
its ability to accommodate gaugino mediated SUSY breaking, on a firmer 
conceptual ground. 

Secondly, given the above discussion, it is possible to localise a SM
field, e.g. a $\bf 2$ zero mode from an $(H,H^c)$ hypermultiplet, at the 
SU(5) brane. Should we be worried by this somewhat counterintuitive 
possibility? The answer is no since, in the limit of large bulk mass,
the lowest-lying mode of the $\bf 3$ (which is also localized at the 
SU(5) brane) becomes massless (cf.~Eq.~(\ref{m4})), so that a full $\bf 5$
emerges.

\section{A three-generation flavour model}\label{fla}

Let us now proceed by using the above tools to construct a 5d SU(5) model
with three generations and see-saw neutrinos which explains all the mass
and mixing hierarchies of the standard model. 

As input we assume a small separation between the scale of the bulk masses
and $M$ (e.g., $m/M\sim 0.1$). This is justified since, on the one hand,
$M$ is the fundamental 
scale of the bulk theory and, on the other hand, $m\sim M$ would imply the 
localization of zero modes on length scales $\sim 1/M$ -- a situation 
outside the realm of our effective field theory approach. Given the 
hierarchy between $M_c$ and $M$, this implies that $ml$ is large (e.g., 
$ml\sim 10$). Such a situation is phenomenologically attractive since
the localization of zero modes is sufficiently strong to 
produce a large fermion mass hierarchy. 

We define the SU(5) gauge sector of our model as explained at the beginning 
of Sect.~\ref{orbbm} and introduce two Higgs doublets $(H_u,H_u^c)$, 
$(H_d,H_d^c)$ as well as three $\bf\bar{5}$'s 
$(\bar{F}_i,\bar{F}_i^c)$ as hypermultiplets in the bulk. Furthermore, we 
distribute the three $\bf 10$'s of the SM at the three distinct locations 
of our model, namely, $T_3$ at the SU(5) brane, $T_2$ in the bulk,
and $T_1$ at the SM brane.  More precisely~\cite{HN,HMR}, for $T_2$ we
have to introduce $(T_2,T_2^c)$ and $(T_2',T_2'^c)$ choosing opposite
$P'$ action between the two, so that a full $\bf 10$ of zero modes
emerges.  For $T_1$ we mean that states with quantum numbers of a full
${\bf 10}$ are located on the SM brane.  (As discussed above the correct
quantum numbers for $T_1$ automatically follow if these states
are understood as localized bulk fields.  $T_3$ can also be
thought of as the limit of a bulk field.) 
The location of fields is shown in Fig.~1.

We allow all Yukawa couplings consistent with gauge symmetry and R parity 
(see, e.g.,~\cite{HN}) at both branes with ${\cal O}(1)$ dimensionless
coefficients. The hierarchical structure of the effective 4d Yukawas will
be entirely due to the different normalization of bulk vs. brane fields and 
to the bulk-mass-driven localization. To begin, let us denote a Yukawa 
coupling between 3 brane superfields by $\lambda$. If one of the three 
fields is replaced by a bulk field with $y$-independent zero mode, the 
effective 4d Yukawa coupling is rescaled according to $\lambda\to\lambda/ 
\sqrt{Ml}$ (following~\cite{AHDDMR} and~\cite{HMROS,hns}). Here the factor 
$M^{-1/2}$ arises because of the mass dimension of the coefficient of the 
original brane-bulk interaction of the 5d theory (the natural scale being 
$M$) and the factor $l^{-1/2}$ comes from the different normalization of the 
kinetic term for brane and bulk fields. If the bulk field has a mass term 
$m$, so that the zero mode is localized as in Eq.~(\ref{zm}), the 
corresponding rescaling reads
\beq
\lambda\to \frac{\lambda}{c(-ml)\sqrt{Ml}}\qquad\mbox{or}\qquad
\lambda\to \frac{\lambda}{c(ml)\sqrt{Ml}}\,
\eeq
depending on whether the original 5d interaction is localized at the SU(5) or 
the SM brane. Here the coefficient function
\beq
c(ml)=\sqrt{\frac{e^{2ml}-1}{2ml}}
\eeq
takes into account the proper normalization of the 5d vs. 4d kinetic terms
and the value of the zero mode at the respective branes. 

Taking $\lambda\sim 1$ for all interactions, introducing bulk masses $m_u$ 
and $m_d$ for the two Higgs hypermultiplets (with $m_ul,\,\,m_dl\gg 1$), 
and keeping all other bulk masses zero for simplicity, we arrive at the 
following Yukawa matrix structure for the two effective 4d interactions 
$H_uT^T\lambda_{TT}T$ and $H_dT^T\lambda_{TF}\bar{F}$:
\beq
\lambda_{TT}=\lambda_t\left(\begin{array}{ccc}
\delta_u & \epsilon\delta_u & 0\\
\epsilon\delta_u & \epsilon^2 & \epsilon\\
0 & \epsilon & 1 
\end{array}\right)\,\,,\qquad\lambda_{TF}=\lambda_b\left(\begin{array}{ccc}
\delta_d & \delta_d & \delta_d\\
\epsilon & \epsilon & \epsilon\\
1 & 1 & 1 
\end{array}\right)\,\,.
\eeq
Here we have used the definitions
\beq
\lambda_t=\sqrt{\frac{2m_u}{M}}\,,\quad
\lambda_b=\frac{1}{\sqrt{Ml}}\sqrt{\frac{2m_d}{M}}\,,\quad
\epsilon=\frac{1}{\sqrt{Ml}}\,,\quad
\delta_u=e^{-m_ul}\,,\quad \delta_d=e^{-m_dl}\,.\label{ma}
\eeq
We recall that we are only attempting to generate the correct hierarchical 
Yukawa structure and that unknown ${\cal O}(1)$ coefficients (including 
complex phases) multiply each of the entries of the above matrices. 

\begin{figure}
\begin{center}
\includegraphics[width=4in]{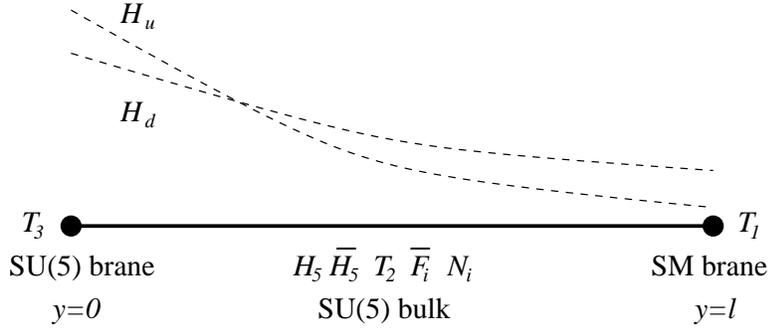}
\end{center}
\caption{The location of the ${\bf 5}$ and ${\bar {\bf 5}}$ Higgs
and 3 generations of matter.  The dotted lines
schematically illustrate the bulk profile of the massless doublet-Higgs
$H_u$ and $H_d$ states.  The $N_i$ are three SM singlet rhd neutrino
states as discussed in Sect.~5.}
\label{fig:flav5d}
\end{figure}

The eigenvalues of the matrices multiplying $\lambda_t$ and $\lambda_b$ in 
Eq.~(\ref{ma}) are $(1,\epsilon^2,\delta_u)$ and $(1,\epsilon,\delta_d)$. 
Successful phenomenology requires the GUT scale relations (see, 
e.g.,~\cite{fk}) $\lambda_t\sim 0.6$, $(\lambda_t\tan\beta)/\lambda_b\sim 
110$, $m_t/m_c\sim\epsilon^{-2}\sim 300$, $m_b/m_s\sim\epsilon^{-1}\sim 30$, 
$m_t/m_u\sim\delta_u^{-1}\sim 10^5$ and $m_b/m_d\sim\delta_d^{-1}\sim 10^3$.
For a moderate value of $\tan\beta\simeq 5$, and up to ${\cal O}(1)$ factors
that depend on the precise brane Yukawa couplings, this set of
flavour hierarchies is realized by taking
\beq 
Ml\simeq 300\,,\qquad m_ul\simeq 11.5\,,\qquad m_dl\simeq 6.9\, .
\label{values}
\eeq
These values are within the favoured parameter range for $Ml$ and for the 
ratio of bulk and brane masses (see~Sect.~\ref{orbbm} and the beginning of 
this section). An illustration of the essential features of our complete 
setup is given in Fig.~1. 

Note that a very similar model is obtained by interchanging the positions 
of SU(5) and SM brane in the setup of Fig.~1. This configuration, where the 
Higgs fields are now peaked at the SM brane, has the advantage that the 
effective 4d Higgs triplet masses do not fall below $M_c$ in the limit of a 
large bulk mass term. Thus, their effect on the precision gauge coupling
unification is guaranteed to remain small.

To derive the resulting structure of the CKM matrix, recall that
$\lambda_{TT}$ and $\lambda_{TF}$ are diagonalized by the bi-unitary
transformations $\lambda_{TT}^{\rm diag}=L^\dagger_T\lambda_{TT}R_{T}$ and
$\lambda_{TF}^{\rm diag}=L^\dagger_F\lambda_{TF}R_F$. (Note that, in our
approach, SU(5) breaking effects in the Yukawa couplings arise only from
unknown ${\cal O}(1)$ coefficients.) Using the fact that, in our model, we
approximately have $\delta_u\sim\delta_d^2\sim\epsilon^4$, the following
structure results:
\beq
V_{\rm CKM}=L^\dagger_TL_F\sim\left(\begin{array}{ccc}
1 & \epsilon & \epsilon^2\\
\epsilon & 1 & \epsilon\\
\epsilon^2 & \epsilon & 1
\end{array}\right)\,.
\eeq
With our choice of $\ep=1/\sqrt{M l}\simeq 1/17$, this compares favourably
with the data as far as 2-3 and 1-3 mixings are concerned.  However, we
underestimate 1-2 mixing by a factor $\sim 4$.  Although we have to admit
that this is arguably the weakest point of our model, we would like to
emphasize that a very modest enhancement of one of the
off-diagonal entries in $\lambda_{TF}$ is sufficient to explain the
large observed Cabibbo angle.

\section{Neutrino masses and mixings}

An important aspect of flavour physics in orbifold GUTs is the generation 
of large neutrino mixing angles and the overall neutrino mass scale. In a 
straightforward approach, one could introduce 3 rhd neutrino singlets $N_i$ 
with ${\cal O}(1)$ Yukawa couplings and Majorana masses $\sim M$ on the SU(5) 
brane. However, given that $M$ tends to be larger than $10^{16}$ GeV and 
the effective 4d Yukawas are suppressed because $H_u$ is a bulk field, the 
resulting light neutrino masses generated by the see-saw mechanism come out
too small. In the following, we discuss two possibilities for obtaining a 
realistic neutrino mass scale in 5d orbifold GUTs, both of which make 
essential use of bulk mass terms.\footnote{The use of bulk singlet states
as rhd neutrinos was previously analysed for large extra dimensions
\cite{AHDDMR}, and mentioned in Ref.\cite{HMROS} in the context
of orbifold GUTs. Our analysis differs from these previous discussions.}

In our first scenario, we introduce three bulk hypermultiplets $(N_i,N^c_i)$, 
which are singlets under SU(5), with parity assignments $(+,+)$ and $(-,-)$ 
for the chiral components $N$ and $N^c$, respectively. In addition, we gauge 
U(1)$_{\chi}$ (named following~\cite{pdg}) in the bulk, where 
${\rm SU}(5)\times {\rm U}(1)_{\chi} 
\subset {\rm SO}(10)$, with orbifold boundary conditions that break 
U(1)$_{\chi}$ at the SM brane. (Specifically this is achieved by $(+,-)$ 
and $(-,+)$ parities for $A_\mu^{\chi}$ and $A_5^{\chi}$, and opposite 
assignments for the gaugino partners. This leaves no U(1)$_{\chi}$ zero 
modes. Note that these assignments require the U(1)$_{\chi}$ gauge coupling 
to be odd under orbifolding.)  Under the bulk U(1)$_{\chi}$, the charges of 
the various states 
are $\chi(T_i) =-1$, $\chi(\bar{F_i}) = 3$, $\chi(H_{u,d})=\pm 2$,
and $\chi(N_i)=-5$, and opposite 
for the conjugate chiral superfield in each hypermultiplet.\footnote{To 
ensure anomaly freedom and the absence Fayet-Iliopoulos terms,
the sum of the charges
of the brane localized fields plus half the sum of the 
charges of bulk fields with even boundary conditions at that brane have to 
vanish~\cite{fi}. Given the $T_i$, $T_i'$, $\bar{F}_i$ and $\bar{F}_i'$ 
of Sect.~4 and the anomaly freedom of SO(10), this is easily 
realized by adding partner hypermultiplets $N_i'$. They will not interfere 
with the neutrino mass generation described below if they are peaked at the 
SM brane.}

At leading order the most general 5d superpotential for the $N_i$'s
consistent with the above symmetries takes the form
(where $L_i$ are the lepton doublets contained in $\bar{F}_i$)
\beq
N^{cT}(\p_5 + m_{N}) N +
H_u L^T \la  N \de(yM) + H_u L^T  \la'  N \de([y-l]M) +
M N^T \kappa  N \de([y-l]M) \,.
\eeq
Here $\la,\la',m_{N}$ and $\kappa$ are $3\times3$ matrices in generation
space.  Note that the $N^c_i$ can not have a similar Majorana mass term 
since they vanish at the SM brane.  Because the $H_u$ Higgs is highly peaked 
towards the SU(5) brane, the $\la'$-term can be neglected. 

Appealing to a complete flavour symmetry of the 5d bulk which acts on 
the ${\bar F}_i$ and $N_i$, we take the matrix $m_N$ to be of the form
$m_N{\bf 1}_3$, where $m_N>0$. In the 
absence of the Majorana mass term, one would find 3 zero modes of the 
superfields $N_i$ with bulk profile $\exp(-m_Ny)$. The Majorana mass couples 
these zero modes and, given the exponential suppression of the bulk profile
at the SM brane and properly normalising the effective 4d kinetic term,
leads to the effective 4d rhd neutrino mass matrix
\beq
M_R \simeq 4 \kappa m_N e^{-2m_{N}l}\,.
\eeq
It is a simple exercise to check that this result also follows from first 
deriving the exact solution of the equations of motion, including the
$N^T \kappa N \de(y-l)$ interaction, and then expanding the expression for 
the light-mode mass-matrix to leading order in $e^{-2m_{N}l}$. 
For moderate values of $m_{N}l$, the mass eigenvalues in $M_R$ are still 
super-heavy, but they are parametrically lighter than $M$. 
Integrating out these modes, the usual see-saw mechanism now generates the 
light Majorana neutrino mass matrix. The relevant term in the low-energy 4d 
superpotential is
\beq
{(\la^T M_R^{-1} \la)_{ij}\over 2(Ml)^3 c(-m_u l)^2 c(-m_N l)^2}
(L_i H_u) (L_j H_u)\,.
\eeq

Because all $N_i$ and $\bar{F}_i$ are treated on an equal footing, the 
resulting form of the light neutrino mass matrix in generation space is 
non-hierarchical,
\beq
{\vev{H_u}^2 (\la^T M_R^{-1} \la)\over (Ml)^3 c(-m_u l)^2 c(-m_N l)^2} \simeq
{m_u l\, v^2\, e^{2m_N l}\over 2M (Ml)^2}
\left(\begin{array}{ccc}
1 & 1 & 1 \\
1 & 1 & 1 \\
1 & 1 & 1 
\end{array}\right) ,
\label{nutex}
\eeq
where, of course, unknown ${\cal O}(1)$ factors multiply the different 
entries, and $v=246\gev$.  As a result the neutrino mixing angles
are naturally large, and 
the super-light neutrino mass-differences are not strongly hierarchical. 
One may in principle be worried about the CHOOZ constraint
$\th_{e3} < 0.16$, but an analysis of the same texture structure
in ``neutrino mass anarchy'' models shows that no particular fine tuning
is necessary for there to be one accidentally small mixing
angle~\cite{anarchy} (see, however,~\cite{wy} for a recent 6d orbifold 
model addressing this issue). Taking the parameters of Eq.(\ref{values})
and assuming $M\simeq 10^{17}\gev$, we find that a reasonable value of 
$m_N l\simeq 6.8$ leads to a phenomenologically viable light neutrino mass 
scale of $m_\nu \sim 0.03\ev$.

A second scenario is even simpler.  We again have $N_i$ and $N_i^c$
chiral superfields, but choose the orbifold action to be $(+,-)$ and
$(-,+)$ respectively.  We further take the superpotential to be of the
form
\beq
N^{cT}(\p_5 + m_{N}) N + H_u L^T \la  N \de(yM) +
H_u L^T \la' N^c \de([y-l]M) \, .
\eeq
Note that this superpotential is not the most general that can written, as 
possible brane-localized masses $M_NNN\de(yM)$ and $M_N'N^c N^c\de([y-l]M)$ 
have been set to zero. (This is technically natural due to supersymmetry.) 
If one is willing to accept this, then the odd bulk mass $m_{N}$ leads to an
exponential suppression of the 4d mass connecting $N$ and $N^c$. Integrating
out this mode then gives an $(LH_u)^2$ operator with a coefficient that can 
easily accommodate the correct light neutrino mass scale. The required 
value of $m_N$ is larger than in our first scenario since the resulting 
exponential factor has to compensate for the weak coupling of the $H_u$ zero
mode at the SM brane. Once again, because of the symmetrical treatment of 
the $N_i$ and $\bar{F}_i$ bulk modes, which is only broken by brane Yukawa 
interactions, large neutrino mixing angles are natural.

\section{Conclusions}

In this letter we have argued that there exists, in the context of a
5-dimensional orbifold SU(5)-GUT model, an appealing explanation of
the observed hierarchical structure of the quark and lepton masses and
mixing angles.  Our model uses only ingredients intrinsic to orbifold
GUT constructions.  These are the existence of branes fixed by the
orbifold action on which gauge and other symmetries can be violated,
and the presence of bulk mass terms for the bulk hypermultiplets.
Our model has the attractive feature that it does not invoke high-scale
Higgs breaking.  Flavour hierarchies arise from two effects: first,
the geometrical suppression of the couplings of bulk fields, as
compared to the couplings of brane fields;  second,
bulk masses leading to partial localization, or 4d mass scale suppression.
Our model provides a simple and concrete demonstration that the
observed flavour hierarchies (dimensionless ratios $\gsim 4$ or larger)
can be explained in geometrical terms within the elegant framework of a
5d orbifold GUT.

Concerning neutrinos, we have shown that there are two attractive
higher-dimen\-sional variations of the traditional see-saw mechanism.
Both take the rhd neutrino states to be modes of SU(5)-singlet bulk
hypermultiplets, $N_i$.
The first involves the gauging of an additional U(1)$_\chi$ in the bulk
which is broken on the SM brane by the orbifold action.  The allowed
masses for the rhd neutrino states are then a large Majorana mass on the 
SM brane and a bulk $NN^c$ mass. The latter leads to a suppression
of the effective 4d Majorana masses of the lightest 4d rhd states and thus
to a suitably enhanced coefficient of the $(LH_u)^2$ operator compared to 
the naive $1/M\simeq (10^{17}\gev)^{-1}$. 
The second model does not involve any additional bulk gauge symmetry or 
brane-localized Majorana mass, but flips the sign of the orbifold action 
to forbid zero modes of $N$ and $N^c$. In this case the bulk mass
suppresses the 4d mass of the lightest KK modes. Both mechanisms naturally
lead to large mixing angles as the bulk structure of the ${\bar F}_i$'s
and $N_i$'s is independent of generation, with this symmetry being only
weakly broken by brane interactions.

\noindent
{\bf Acknowledgements}:
We are grateful to Yasunori Nomura and, particularly, Riccardo Rattazzi 
and for helpful conversations.

\end{document}